\documentclass[superscriptaddress,secnumarabic,
amssymb,amsmath,nobibnotes,aps,prd,showkeys,showpacs,nofootinbib]{revtex4}%
\usepackage{graphicx}
\usepackage{epsf}
\usepackage{bm}
\usepackage{amsmath}
\usepackage{amsfonts}
\usepackage{amssymb}
\usepackage{epstopdf}
\usepackage{subfigure}
\usepackage{natbib}
\usepackage{color}%
\usepackage{hyperref}
\hypersetup{dvips,dvipdfm,linktoc=page,colorlinks=true,linkcolor=blue,citecolor=red,filecolor=magenta,urlcolor=magenta,bookmarks=true}
\providecommand{\U}[1]{\protect\rule{.1in}{.1in}}
\newcommand{\be}{\begin{equation}}
\newcommand{\ee}{\end{equation}}

\newcommand{\mincir}{\raise
-3.truept\hbox{\rlap{\hbox{$\sim$}}\raise4.truept\hbox{$<$}\ }}
\newcommand{\magcir}{\raise
-3.truept\hbox{\rlap{\hbox{$\sim$}}\raise4.truept\hbox{$>$}\ }}

\begin{document}

\title{Phase Space Analysis and Thermodynamics of Interacting Umami Chaplygin gas in FRW Universe}
\author{Sujay Kr. Biswas}
\email{sujaymathju@gmail.com}
\affiliation{Department of Mathematics, University of North Bengal, Raja Rammohunpur, Darjeeling-734013, West Bengal, India.}

\author{Atreyee Biswas}
\email{atreyee11@gmail.com}
\affiliation{Department of Applied Science, Maulana Abul Kalam Azad University of Technology, Haringhata, Nadia- 721249, West Bengal, India}
\keywords{Irreversible thermodynamics; GSLT; Umami Chaplygin gas, Interaction, Dynamical system, Phase space, Stability}
\pacs{95.36.+x, 95.35.+d, 98.80.-k, 98.80.Cq.}
\begin{abstract}
In this work interacting Umami Chaplygin gas has been studied in flat FRW model of universe in context of it's thermodynamic and dynamical behaviour. In particular, considering Umami fluid as dark energy interacting with dark matter, irreversible thermodynamics has been studied both for apparent and event horizon as bounding horizon in two separate cases. Also the model has been investigated in purview of dynamical systems analysis by converting the cosmological evolution equations to an autonomous system of ordinary differential equations. With some restrictions on model parameter $\omega$ and coupling parameter $\lambda$, some cosmologically interesting critical points describing late time accelerated evolution of the universe attracted by cosmological constant and accelerated scaling attractor in quintessence era have been found to alleviate coincidence problem.
\end{abstract}
\maketitle
\section{Introduction}
Recent observations like Supernovae Ia (SNeIa) \cite{Riess1}, cosmic microwave background
(CMB)\cite{Bennet2003} , large scale structure (LSS) \cite{Hawkins2003,Tegmark2006,Cole2005} etc indicate that our universe is now going through an accelerating phase. This invokes scientists to believe in the existence of an exotic matter called dark energy (DE) whose negative pressure is thought to drive the late time accelerating expansion of universe. In fact, SNeIa observations have shown that about $70\%$ of present energy budget of universe is contributed by DE. The most obvious choice for DE is the cosmological constant($\Lambda$CDM model) with constant equation of state (EoS). But it suffers from some severe drawbacks like coincidence and fine tuning problem. To avoid these problems alternatively some perfect fluid models have been proposed in several research papers. One can see the review paper by K. Bamba et al \cite{Bamba2012} in this regard.  Among these models, Chaplygin gas type cosmology turns out to be very promising one. This type of models behaves as dust in the early stage of universe while at later stage they can play the role of DE, i.e, they serve to be the possible unification of dark energy and dark matter (DM) and therefore belong to the class of unified model of DE. The original Chaplygin gas model was proposed by Kamenshchik, Moschella, and Pasquier \cite{Kamenschick2001} with EoS
\begin{eqnarray*}
  P = -\frac{B}{\rho}
\end{eqnarray*}
Here $P$ and $\rho$ are the pressure and density of the fluid respectively and $B$ is a constant.
Later Bento et al.\cite{Bento2002} generalized the EoS to
\begin{eqnarray*}
P =-\frac{B}{\rho^{\alpha}}
\end{eqnarray*}
where $\alpha$($0\leq \alpha\leq 1$) generalizes from 1 to any arbitrary constant and this generalized model was called the generalized Chaplygin gas (GCG) model. It can be seen that the GCG almost coincide with dust $(P = 0)$ at high energy density which completely does not go with our Universe. Therefore, modified Chaplygin gas (MCG) was introduced with the following equation of state
\begin{eqnarray*}
P = A\rho-\frac{B}{\rho^{\alpha}}
\end{eqnarray*}
where $A$ is a positive constant.This model is more appropriate choice to have constant negative pressure at low energy density and high pressure at high energy density. The special case of $A = \frac{1}{3}$  is the best fitted value to describe evolution of the universe from radiation regime to the CDM regime. Later many variation of this model such as extended chaplygin (ECG)\cite{Pourhassan2014}, variable chaplygin gas \cite{Guo2007} have been introduced. A lot of research works have been done on chaplygin class of fluids in cosmological context. J. C. Fabris et al \cite{Fabris2002} studied the fate of density perturbations in an Universe dominate by Chaplygin gas where they cosidered it as a Newtonian fluid and therefore performed a newtonian analysis. Their study had shown an initial phase of growing perturbations, with the the same rate as in dust case of the cosmological standard model, from which it follows decreasing oscillations, which asymptotically go to zero. Therefore from this result it follows that though newtonian background model is very different from the relativistic one, perturbative newtonian analysis is consistent with relativistic considerations.  Also it was revealed that Chaplygin gas, in spite of presenting negative pressure, is stable at small scale, in opposition to what happens in general with perfect fluids with negative pressure. This is due to the fact that the sound velocity in the Chaplygin gas is positive. In an another work J. C. Fabris et al \cite{Souza2002} had shown that Chaplygin gas can not only serve as an unification of dark energy-dark matter but also can be a good candidate of DE models.  They analyzed some consequences of this scenario using type Ia supernovae data (SNe Ia). Their results indicate that a cosmology completely dominated by the Chaplygin gas is favored in comparison to $\Lambda$CDM models. The trajectories of statefinder parameters in Chaplygin class of models were studied in Ref. \cite{Gorini2003}  while constraints involving Cosmic Microwave Background (CMB) data, Fanaroff-Ryley type IIb radio galaxies and X-ray data from galaxy clusters, have also been extensively discussed by many authors either as a dark energy or in the unified dark matter-energy (UDME) picture \cite{Silva2003,Amendola2003}. Inflationary scenario in the presence of generalized Chaplygin gas (GCG) was studied in a recent work by B. R. Dinda et al \cite{Dinda2014}. They had shown that in Einstein gravity, GCG is not a suitable candidate for inflation; but in a five-dimensional brane-world scenario, it can work as a viable inflationary model. M. R. Setare \cite{Setare2007} analyzed a correspondence between the holographic dark energy density and Chaplygin gas energy density in FRW universe and reconstructed the potential and the dynamics of the scalar field which describe the Chaplygin cosmology. S. Li et al \cite{Li2009} studied dynamical evolution of universe considering MCG interacting with cold dark matter. They had shown that there exists a stable scaling solution, which provides the possibility of alleviating the coincidence problem. The growth of large scale structure was studied by $T.Multam\ddot{a}ki$ \cite{Manera2004} in a universe containing both cold dark matter (CDM) and a generalized Chaplygin gas (GCG). Here, the GCG only contributed to the background evolution of the universe, while the CDM component formed structures showing both linear and as well as nonlinear growth. Though this model passed the standard cosmological distance test without the need of a cosmological constant, but it was proved to be severely constrained by current observations of large scale structure. Fate of universe for Chaplygin gas class models in context of thermodynamics have also been studied by several authors. S. Chakraborty et al \cite{Chakraborty2019} studied validity of generalized second law of thermodynamics(GSLT) for different Chaplygin gas models and had shown that though some models obey GSLT on apparent horizon, but in case of event horizon its validity depends on the choice of free parameters in the respective models. A Biswas \cite{Biswas2018} in context of standard Eckart theory of irrevrsible thermodynamics had shown that GSLT remains valid for an interacting MCG model on both apparent and event horizon subject to some restrictions on model parameters and these restrictions support observational results.  Recently a new generalization of Chaplygin gas model namely Umami Chaplygin gas, has been proposed by R. Lazkoz et al \cite{Lazkoz2019} with EoS
\begin{equation}
P=-\frac{\rho}{\frac{1}{|\omega|}+\frac{\rho^2}{|A|}}\label{1}
\end{equation}
where $\omega$ and $A$ are real constants.The parameters $\omega$ and $A$ play vital role in describing the nature of DE equation of state at low energy as well as high energy density limit. In particular, in high energy limit one obtains $P \longrightarrow -\frac{|A|}{\rho}$ and in the low enegy limit $
P\longrightarrow -|\omega| \rho$ which describe various DE components accordingly \cite{Lazkoz2019}.
It was shown in \cite{Lazkoz2019} that unlike previous versions of Chaplygin gas models, Umami
Chaplygin gas can not only unify dark energy and dark matter but also can unify
dark fluid with baryonic matter, i.e it can contribute the total matter distribution of
the universe. Therefore it is interesting to study this new variation of Chaplygin gas
model as an alternative of $\Lambda$CDM model.\\

Now, the nature of DE and DM are completely unknown to us and an interaction among them can not be overruled. Rather, a proper interaction which is favored by observational data can give a possible mechanism to alleviate the coincidence problem. Also, proper choice of the interaction term may influence the perturbation dynamics and affect the lowest multipoles of the CMB
spectrum \cite{Wang2005,Wang2007}. In fact, in an analysis \cite{Gong2007,Ohta2007} of the supernova data together with CMB and large-scale structure revealed such interaction from expansion
history of the universe. In an interacting scenario, critical points (extracted from autonomous system) which are represented by the late time scaling attractor with \cite{C.G.Bohmer2008,C.Wetterich1995,L.Amendola1999,Sujay2017}
	\begin{equation}\label{coincidence}
		r=\frac{\Omega_{m}}{\Omega_{d}} \approx O (1)~~~~~~~ \mbox{with}~~~~~~~ \omega_{eff}<-\frac{1}{3}
	\end{equation}
 and correspond to accelerating universe, can solve the coincidence problem by fine tuning the model parameters without fine tuning the initial conditions. Moreover, from thermodynamic point of view the interaction between DE and DM has been studied in ref.\cite{Wang2008}  where DE was considered as perfect fluid with a well-defined temperature and it was shown that at present epoch the energy flow should be from DE to DM for the validity of the second law of thermodynamics \cite{Callen1960}. Therefore it is always interesting to study interacting dark energy models from dynamical as well as from thermodynamic point of views. Ineracting DE models have been studied by several authours in Refs. \cite{M.Khurshudyan2015,Yuri.L.Bolotin2014,Odintsov2017,Oikonomou2018,N.Tamanini2015,Xi-ming Chen2009,T.Harko2013,Wang2016,Fang1,Odintsov2018b,Landim22016,Odintsov2018a,Kleidis2018,S.Kr.Biswas2015a,S.Kr.Biswas2015b,Copeland1,Leon1,Bahamonde2018,S.Li2009,J.Bhadra2011,C. Ranjit2014}.\\
In this work, dynamical as well as thermodynamical behaviour of the interacting Umami Chaplygin gas model has been investigated to achieve the overall qualitative evolutionary picture of the universe. In particular, in this paper, we have considered Umami chaplygin gas playing the role of DE with any perfect fluid model equation of state $\gamma_{d}$ (which can be represented as quintessence, cosmological constant or phantom fluid according to parameters restrictions) interacting with DM (considered as any perfect fluid with barotropic equation of state $\gamma_{m}$). For dynamical analysis, cosmological evolution equations are converted into an autonomous system of ordinary differential equations (ODEs) by using dimensionless variables normalized over Hubble scale and then linear stability theory has been applied to find the nature of critical points, i.e, whether they can describe late time accelerated evolution of the universe or late time accelerated scaling solution attracted in quintessence era to alleviate the coincidence problem.\\

On the other hand, for the study of thermodynamics of Universe, it should be noted that after the revolutionary discovery by S.Hawking \cite{Hawking1975} that black hole behaves as a blackbody, it was established that there exists a beautiful connection between black hole mechanics and thermodynamics - surface gravity and area of black hole horizon playing the role of temperature and entropy of black hole.
In particular, black hole horizon, is associated with thermodynamic properties - the Hawking temperature $T$ and entropy $S$ with the following relations:\\

$T = \frac{\kappa}{2\pi}$  and $S=\frac{A}{4G}$\\

where $\kappa$ is surface gravity of blackhole horizon and $A$ being the area of horizon. Here $G$ is Newton's constant. In space time dynamics, heat is defined as energy that flows across a causal horizon. It can be felt via the gravitational field it generates, but its particular form or nature is unobservable from outside the horizon and this horizon need not be only a black hole horizon. In 1995 T. Jacobson \cite{Jacobson1995} derived Einstein’s equation as an equation of state from the proportionality of entropy and horizon area together with the fundamental Clausius relation $\delta Q=T dS$ connecting heat, entropy and temperature for all local accelerating horizon (local Rindler causal horizons) with $\delta Q$ and $T$ interpreted as the energy flux and Unruh temperature seen by an accelerated observer just inside the horizon. On the other hand, T. Padmanabhan \cite{Padmanabhan2002} in 2002 derived first law of thermodynamics on the apparent horizon starting from Einstein’s field equations for a general static spherically symmetric space time. These deductions strongly suggest that the association of entropy and temperature with a horizon of universe is quite fundamental and actually connected
with the dynamics of gravitational field. Just as for black hole, in a cosmological model
with cosmological constant (de Sitter space) there also exists Hawking temperature and
entropy that can be associated with cosmological event horizon. The thermodynamics
in de-Sitter space was first studied by Gibbons and Hawking \cite{Gibbon1977} and it has been extended
to the quasi de-Sitter space \cite{Frolov2003,Pollock1989}.
In recent years black hole thermodynamics has been extended to dark energy cosmology \cite{Wang2005,Gong2006,Gong2007,Bousso2005,Zhang2008}. In 2006 in an attempt to reproduce Einstein's equations in f(R) gravity, Eling et al\cite{Eling2006} noticed that a generalization to a non-equilibrium thermodynamical setting is necessary and therefore, in order to recover the f(R) gravity field equations from first law of thermodynamics, they modified the equilibrium entropy balance relation of the system by adding an extra term called internal entropy production term leading to non-equilibrium thermodynamic prescription. Also,from realistic point of view universal thermodynamics should be irreversible instead of reversible in nature as normally, the process of energy flux crossing the horizon is irreversible and non-equilibrium thermodynamics should be taken into account. In fact, the evolution of the universe contains a sequence of important dissipative processes like Nucleosynthesis,Decoupling of neutrinos from cosmic plasma, Decoupling of photons from matter during recombination era etc and one needs non-equilibrium thermodynamic treatment to explain these processes. At least the effect of heat conduction was not negligible during evolution of universe at early stages and since the matter of universe did not attain thermal equilibrium at early stages, there should have been heat flow during certain stages of evolution. Vanleeumen et al.\cite{Vanleeuwen1975},Novello and Reboucas \cite{Reboucas1982}, Coley and Tupper \cite{Coley1983,Coley1984} Sviestins \cite{Sviestins1985}, Mukherjee \cite{Mukherjee1986}, Banerjee and Sanyal \cite{Banerjee1988}, Deng \cite{Deng1989} all have studied the cosmological models with heat conduction.
Gong et al \cite{Gong2007} studied non-equilibrium thermodynamics for  the universe bounded by apparent horizon with DE in the form of perfect fluid with constant EoS. It was shown there that the radius of the apparent horizon increases depending on the EoS parameter of DE as well as on the  parameter contributing to non-equilibrium thermodynamics. In this paper the work of Gong et al \cite{Gong2007} has been extended to study universal thermodynamics considering Umami Chaplygin gas model of DE for not only apparent horizon as bounding horizon but also for event horizon.\\
The paper is organized as follows:
Section 2 is deals with the study of dynamical stability analysis of Universe containing interacting Umami Chaplygin gas while in section 3 it's thermodynamic behaviour has been investigated in context of standard Eckart theory of irreversible thermodynamics. Finally, Section 4 is devoted for the summary and some concluding remarks.

\section{Dynamical analysis of the model:}
\label{phase_space_auto}

In this section, we shall study our model in the light of dynamical systems analysis. First, we convert the cosmological evolution equations into an autonomous system of ordinary differential equations by suitable transformation of variables. Then, we shall study this system in a compact phase-space by considering the dimension-less dynamical variables which should be undertaken normalized over Hubble scale.

\subsection{ Formulation as an autonomous system}
\label{formation_autonomous}
In this work, we consider the universe to be spatially flat and consists of two interacting fluid - DE and DM. Then The Friedmann equation and acceleration equation are  $\left(\text{using natural units}~  8\pi G=\frac{8 \pi}{m_{PL}^{2}}=\hbar=c=1  \right)$
\begin{equation}\label{Friedmann}
	H^{2} = { \frac{1}{3}}(\rho_d +\rho_m)
\end{equation}
and
\begin{equation}\label{Raychaudhuri}
 \dot{H} = -\frac{1}{2} (\rho_{d}+p_{d}+\rho_{m}(1+\gamma_{m})),
\end{equation}
where, $H$ is the Hubble parameter defines the expansion rate of the universe, expressed as $H=\frac{\dot{a}}{a}$ in terms of scale factor $a(t)$. An over dot indicates the derivative with respect to time $t$. It should be mentioned that Umami chaplygin gas with its energy density $\rho_{d}$ and pressure $p_d$ (as a non-linear function of $\rho_{d}$) plays a role of dark energy candidate which satisfies equation of the state parameter (as in Eqn (\ref{1})):\\
$p_d=-\frac{\rho_{d}}{\frac{1}{|\omega|}+\frac{\rho^{2}_{d}}{|A|}}$. On the other hand, dark matter is considered to be in the form of any perfect fluid with barotropic equation of state $\gamma_{m}=\frac{p_m}{\rho_{m}}$, $p_m$ and  $\rho_{m}$ being the pressure and  enegy density of dark matter. The equation of state parameter $\gamma_{m}$ for dark matter takes any value in between $[0,1]$. The dark matter barotropic fluid behaves as dust for $\gamma_{m}=0$ and as radiation for $\frac{1}{3}$. The variable equation of state parameter for Umami chaplygin gas is denoted by $\gamma_{d}=\frac{p_d}{\rho_{d}}$ and in the current model (using Eqn. (\ref{1})), it is given by
\begin{equation}\label{EoS}
	\gamma_{d}\equiv \frac{p_d}{\rho_{d}}=-\frac{1}{\frac{1}{|\omega|}+\frac{\rho^{2}_{d}}{|A|}}
\end{equation}
It obeys the nature of any perfect fluid model for any real value of the model parameters $A$ and $\omega$.\\
Energy conservation equations for individual fluid component are

\begin{equation}\label{continuity-matter}
	\dot{\rho_{m}} +3H\rho_{m}(1+\gamma_{m})=Q
\end{equation}
and
\begin{equation}\label{continuity-DE}
	\dot{\rho_{d}} +3H(\rho_{d}+p_{d})=-Q.
\end{equation}

Here $Q$ is is the interaction term. It is to be noted that most studies on interaction between DE and DM are chosen depending either on the assumption of interacting field from the out set or on phenomenological requirement. Now, it is interesting to investigate effects of different interaction terms between DE and DM on evolution of universe. In particular, those forms of $Q$ which contribute in the mechanism of alleviating coincidence problem, are quite promising ones. Keeping this motivation in mind and in view of continuity equation (\ref{continuity-matter}) and (\ref{continuity-DE}), one should assume $Q$ to be a function of energy densities multiplied by a factor having dimension inverse of time and Hubble parameter is a natural choice for it. Therefore, $Q$ can take any of the following forms:\\
(i)~ $Q=Q\left(H\rho_m\right)$ \cite{Amendola2001} ,\\
~ (ii)~ $Q=Q\left(H\rho_d\right)$ \cite{Pavon2009},\\
(iii) ~$Q=Q\left(H\left(\rho_m+\rho_d\right)\right)$ \cite{Wang2005}.\\
In this current work for simplicity we have considered the second choice - to be specific\\

\begin{eqnarray}\label{Interaction}
	Q=3\lambda H\rho_{d}
\end{eqnarray}

where $\lambda$ is assumed to be small. Moreover, we have considered $\lambda$ to be positive indicating energy flow from DE to DM, because in \cite{Sen2008} it was argued that negative coupling fails to alleviate coincidence problem and it does not obey second law of thermodynamics also. However, this phenomenologically chosen interaction term is compatible with several observations like SNIa, CMB,large-scale structure, H(z) and age constraints, and recently in galaxy clusters \cite{Wang2006,Wang2007,Gong2007,Bertolami2009,Abdalla2009}. Also, this choice is effective to alleviate the coincidence problem \cite{Steinhardt1997,Zimdahl2001,Chimento2003,Chimento2006,Campo2006}.\\\\
Now, in order to get a qualitative idea of the evolution of the universe we study the phase space analysis by considering the following dimensionless variables \cite{M.Khurshudyan2015,Sujay2017}:

\begin{equation}\label{Variables}
    x=\frac{\rho_{d}}{3H^{2}},~~~~~~  y= \frac{p_{d}}{3H^{2}}.
\end{equation}
These choices of dnamical variables have also been adopted in studying various form of equation of states of dark energy in dynamical analysis (see ref. (\cite{Aljaf})). Using the variables (\ref{Variables}), the Friedmann equation (\ref{Friedmann}) puts a constraint on the dimensionless density parameter as $\Omega_{m}\equiv \frac{\rho_{m}}{3H^2}=1-x$. This constraint together with the energy conditions imposes a restriction on variable $x$ as: $0\leq x\leq1$. Now, the constraint equation, namely, $\Omega_{m}+x=1$ implies that the evolution equation of one variable ($x$ or $\Omega_{m}$) is obtained by substituting the other.  Also, by using equation of state parameter in (\ref{EoS}), the dimensionless variable $y$ ($y=y(x,H)$)-- in terms $H$ (Hubble parameter) and $x$ (variable)  takes the form:
\begin{equation}
	y=-\frac{x}{\frac{1}{|\omega|}+\frac{9x^2 H^4}{|A|}},
\end{equation}		
which by inverse operation gives the Hubble parameter ($H=H(x,y)$) in terms of dynamical variables $x,~y$ and the model parameters $A$ and $\omega$ as follows:
\begin{equation}\label{H}
	H^4 =-\frac{|A|}{9x^2 y}\left(x+\frac{y}{|\omega|}\right).
\end{equation}
Now with the  help of above equation and eliminating the $H$ term, the dynamical variables $x$ and $y$ lead to the following 2D autonomous system of ODEs for the interaction term (\ref{Interaction}):.
\begin{eqnarray}
\begin{split}
   \frac{dx}{dN}& = 3(x - 1)(-\gamma_{m} x + y) -3\lambda x,& \\
    \frac{dy}{dN}&=y \left[\left(-3x - 3y - 3\lambda x \right)\left(\frac{1}{x} -\frac{ 2(x + \frac{y}{|\omega|})}{x^2}\right) + 3 + 3y + 3\gamma_m (1 - x)\right] &~~\label{autonomous}
\end{split}
\end{eqnarray}

Where $N=ln~a$ is the lapse time, the independent variable here,  called e-folding parameter. It should be noted that the autonomous system does not involve the parameter $A$ but it appears in the expression of Hubble parameter in (\ref{H}). As a result, the cosmological parameters like deceleration parameter, density parameters obtained at critical points do not involve explicitly the model parameter $A$. So, we can say that the parameter $A$ does not affect the qualitative dynamics of the evolution of the universe. Now, physical parameters can be expressed in terms of dynamical variables $x$ and $y$ ( the coordinates of critical points) as follows:\\
The density parameter for dark matter can take the form
\begin{equation}
	\Omega_{m} =1-x,
\end{equation}
and the density parameter for dark enegy i.e, for Umami chaplygin gas takes the value:
\begin{equation}
  \Omega_{d} =x,
\end{equation}
The equation of state parameter for Umami Chaplygin gas reads as:
\begin{equation}
	\gamma_{d} = \frac{p_{d}}{\rho_{d}} =\frac{y}{x},
\end{equation}
[which is a non-constant parameter leading to the condition to be a dark energy
		$\gamma_{d}=\frac{y}{x}<0$.]
The deceleration parameter can be expressed (using acceleration equation (\ref{Raychaudhuri})) in the form
\begin{equation}\label{acceleration}
	q= -1-\frac{\dot{H}}{H^{2}}= -1 + \frac{3}{2} \left[1+y+\gamma_{m}(1-x) \right],
\end{equation}
and the effective equation of state parameter for the model is
\begin{equation}\label{effective-eos}
	\omega_{eff}=y+\gamma_{m}(1-x).
\end{equation}
Now, one obtains conditions for acceleration either for $q<0$ or $\omega_{eff}<-\frac{1}{3}$. For the energy condition $0\leq\Omega_{d}\leq1$, the phase space becomes bounded in the following region
\begin{equation}\label{phase boundary}
	0\leq x \leq 1.
\end{equation}

Using the physical parameters, we shall now study the phase space analysis for the system (\ref{autonomous}) with the interaction term (\ref{Interaction}).

\subsection{Phase space analysis:}

There are three critical points obtained from the autonomous system. The critical points and cosmological parameters obtained are shown in tabular form in the Table \ref{parameters}.\\

{\bf The critical point} $A=(\frac{\gamma_{m}-\lambda}{\gamma_m},0)$ exists for the parameter restrictions $ 0\leq \lambda \leq \gamma_{m} $ in the phase space. It is the point with combination of both DE and DM in its evolution. Here, DE behaves as dust (since $\gamma_{d}=0$) and DM takes the form of any perfect fluid except dust ($\gamma_{m}\neq 0$). The point is always decelerating (since $q=\frac{1}{2}+\frac{3\lambda}{2}>0$) in nature due to its existence criteria. For $\lambda\longrightarrow 0$, DE dominates over dark matter and also evolution of the universe is driven by dust ($\omega_{eff} \longrightarrow 0$). On the other hand, for $\lambda\longrightarrow \gamma_{m}$ dark matter(DM) dominates over DE ($\Omega_{d}\longrightarrow 1$) and the expansion is always deceleraing.  These are shown in Table \ref{parameters}.
The eigenvalues of the linearized Jacobian matrix for the point are:
\begin{itemize}
	\item $\mu_{1}=6+6\lambda$,
	\item $\mu_{2}=-3\gamma_{m}+3\lambda.$
\end{itemize}
The point is hyperbolic type since all the eigenvalues are non-zero. It will be non hyperbolic in nature when $\lambda=\gamma_{m}$.
The stability of the point can be investigated by the sign of eigenvalues. Since the critical point A having one eigenvalue namely, $\mu_{1}$ is always positive, the point behaves as a saddle (unstable) point in phase space. \\

{\bf The critical point}

$B=\left(\frac{\gamma_{m}-\lambda+|\omega|}{\gamma_{m}+|\omega|},-\frac{(\gamma_{m}-\lambda+|\omega|)|\omega|}{\gamma_{m}+|\omega|}\right)$ exists for $0\leq \lambda \leq \gamma_{m}+|\omega|$, where $\gamma_{m}$ takes any value in the range $[0,1]$. The point is also a solution with combination of both DE and DM and the ratio of two energy densities is: $r=\frac{\Omega_{m}}{\Omega_{d}}=\frac{\lambda}{\gamma_{m}-\lambda+|\omega|}$.
It should be mentioned that the Dark energy(DE) (with equation of state $\gamma_{d}$) can mimic quintessence, cosmological constant, and phantom fluid respectively according as $\frac{1}{3}<|\omega|<1$,~~$|\omega|=1$, and $|\omega|>1$.
From the Table \ref{parameters}, we observe that acceleration is possible either for\\
(i) $\lambda<-\frac{1}{3}$ \\
 or for\\
(ii) $ \lambda \geq -\frac{1}{3}$ with  $\left\{\omega <\frac{1}{3} (-3 \lambda -1)~~ \mbox{or}~~  \omega >\frac{1}{3} (3 \lambda +1)\right\}$.
A Phantom era (with $\omega_{eff}<-1$) can be achieved in the parameter space of\\
  (i) $\lambda <-1$~~ or\\
  (ii) $\lambda \geq -1~~ \mbox{with}~~ \omega <-\lambda -1 ~~\mbox{or}~~  \omega >\lambda +1$.\\
The critical point B becomes completely Umami Chaplygin gas (DE) dominated solution ($\Omega_{d}\longrightarrow 1$) for $\lambda\longrightarrow 0$ and as a result, an accelerating universe is always appeared near the point. On the other hand, for $\lambda\longrightarrow \gamma_{m}+ |\omega|$, the point becomes a DM dominated solution ($\Omega_{m}\longrightarrow 1$) during it's evolution which implies that there will be deccelerating phase of the universe near the point (because for this case, $q>0$ always)(see Table \ref{parameters}). Interestingly, the critical point has scaling nature in the evolution satisfying Eqn. (\ref{coincidence}).
The eigenvalues of linearized Jacobian matrix for the critical point B are
\begin{itemize}
	\item $\mu_{1}=-6-6\lambda+6|\omega|$,
	\item $\mu_{2}=-3\gamma_{m}+3\lambda-3|\omega|.$
\end{itemize}
The point will be a stable solution (when $\mu_{1}<0,~\mu_{2}<0$) in phase space under certain conditions in the following:\\
(1). $\lambda\geq \gamma_{m},~~ -\gamma_{m}+\lambda<\omega<\lambda+1$, ~~~~or\\
(2). $ 0\leq \lambda<\gamma_{m}, ~~ 0\leq\omega<\lambda+1 $,~~~~~~or\\
(3). $ \lambda\geq \gamma_{m},~~  -\lambda-1<\omega<\gamma_{m}-\lambda$,~~~~or\\
(4). $ 0\leq \lambda<\gamma_{m}, ~~ -\lambda-1\leq\omega<0 $.\\
From the above analysis, one can conclude that late time scaling solution B behaves as a stable attractor in the quinessence era (i.e., $-1<\omega_{eff}<-\frac{1}{3}$) satisfying Eqn.(\ref{coincidence}) under the parameter region $|\omega|<1+\lambda$. Also, the point B evolves in the phantom regime in the parameter space $|\omega|\geq 1+\lambda$, but it corresponds to an unstable (saddle like) solution and obviuosly, it can not become a late time solution there.\\

Finally, {\bf the critical point}

$C=\left(\frac{\gamma_{m}+1}{\gamma_{m}+\lambda+1},-\frac{(\gamma_{m}+1)(\lambda+1)}{\gamma_{m}+\lambda+1}\right)$ exists for $ \lambda \geq 0$ and for all values of $\gamma_{m}$ (in $[0,1]$) and  model parameter $\omega$ in the phase space. Here, the Umami chaplygin gas (DE) describes phantom fluid (since $\gamma_{d}<-1$ see Table \ref{parameters}) always. This point is also a combination of both the Umami Chaplygin gas(DE) and DM with the ratio of two energy densities: $r=\frac{\Omega_{m}}{\Omega_{d}}=\frac{\lambda}{\gamma_{m}+1}$. For $\lambda\longrightarrow 0$, DE behaves as a cosmological constant like fluid and it dominates over DM. The critical point C is always accelerating in nature (since $q=-1$) and it represents the late time de Sitter like solution ($\omega_{eff}=-1$). \\
The eigenvalues of linearized jacobian matrix are:\\
\begin{itemize}
	\item $\mu_{1}=\frac{6(\lambda+1)(\lambda-|\omega|+1)}{|\omega|}$,~~
	\item $\mu_{2}=-3-3\gamma_{m}.$
\end{itemize}
Since all the eigenvalues are non-zero, the critical point C is hyperbolic in nature  and it is stable either for\\
(1) $0\leq \lambda<\omega-1$, when $\omega>1$\\
or for\\
(2) $ 0\leq \lambda<-\omega-1  $, when $\omega<-1.$\\
In summary, when $|\omega|>1+\lambda$, the point exhibits the late time scaling attractor driven by cosmological constant. Moreover, the point corresponds to late time de Sitter like solution in the above parameter space.

\begin{table}[tbp] \centering
	\caption{Critical Points, corresponding cosmological parameters and stability of critical points for the interaction model (\ref{Interaction}) are presented, where $0 \le \gamma_{m} \le 1$.}%
	\begin{tabular}
		[c]{cccccccc}\hline\hline
		\textbf{Critical Points}&$(\mathbf{x,y})$& $\mathbf{\gamma_{d}}$ &
		$\mathbf{\Omega_{m}}$ & $\mathbf\Omega_{d}$ & $\mathbf{\omega_{eff}} $  &
		\textbf{q} & \textbf{Stability}\\\hline
		$A  $ & $( \frac{\gamma_{m}-\lambda}{\gamma_{m}},0 )$&  $ 0 $ &
		$ \frac{\lambda}{\gamma_{m}} $ & $\frac{\gamma_{m}-\lambda}{\gamma_{m}}$ & $ \lambda $ & $ \frac{1}{2}+\frac{3\lambda}{2}  $ & \it{ Always saddle, since} \\&&&&&&&$0 \le \lambda \le \gamma_{m}.$\\
		$B  $ & $\left(\frac{\gamma_{m}-\lambda+|\omega|}{\gamma_{m}+|\omega|},-\frac{(\gamma_{m}-\lambda+|\omega|)|\omega|}{\gamma_{m}+|\omega|}\right)$ & $ - |\omega| $ &
		$ \frac{\lambda}{\gamma_{m}+|\omega|} $ & $ \frac{\gamma_{m}-\lambda+|\omega|}{\gamma_{m}+|\omega|} $ & $ -|\omega|+\lambda $ & $ \frac{1+3(-|\omega|+\lambda)}{2} $ & \it{Stable node for}:\\ &&&&&&& $\lambda\geq \gamma_{m},~~ -\gamma_{m}+\lambda<\omega<\lambda+1,$~ or\\&&&&&&&$0\leq \lambda<\gamma_{m}, ~~ 0\leq\omega<\lambda+1$, ~or\\&&&&&&&$\lambda\geq \gamma_{m},~~  -\lambda-1<\omega<\gamma_{m}-\lambda$,~ or\\&&&&&&&$0\leq \lambda<\gamma_{m}, ~~ -\lambda-1\leq\omega<0$   \\&&&&&&& \it{Saddle node for}:\\ &&&&&&& $|\omega|>1+\lambda, ~~ \lambda \ge 0.$   \\
		$C  $ & $ \left(\frac{\gamma_{m}+1}{\gamma_{m}+\lambda+1},-\frac{(\gamma_{m}+1)(\lambda+1)}{\gamma_{m}+\lambda+1}\right)$ & $ -\lambda-1  $ &
		$ \frac{\lambda}{\lambda+\gamma_{m}+1} $ & $ \frac{\gamma_{m}+1}{\lambda+\gamma_{m}+1} $ & $ -1 $ & $-1$ & \it{Stable node for}:\\ &&&&&&& $0 \le \lambda<|\omega|-1,~~|\omega|>1.$\\&&&&&&& \it{Saddle node for}:\\&&&&&&& $|\omega|\ge 1,~~ \lambda>-1+|\omega|$,~ or\\&&&&&&& $|\omega|<1,~~\lambda \ge0.$
		\\\hline\hline
	\end{tabular}
	\label{parameters}
\end{table}%
%

\begin{figure}
	\centering
	\subfigure[]{%
		\includegraphics[width=8cm,height=6cm]{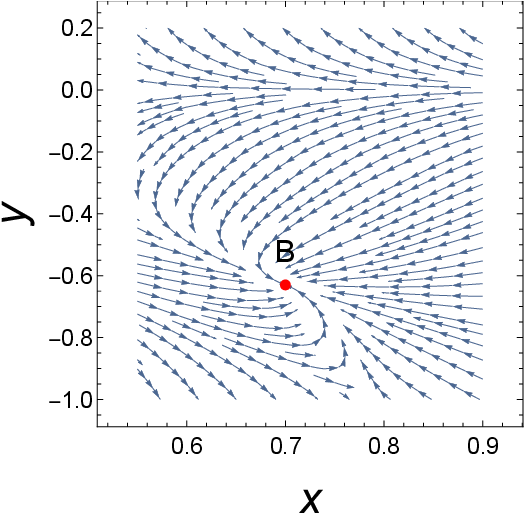}\label{fig:B_Obsv}}
	\qquad
	\subfigure[]{%
		\includegraphics[width=8cm,height=6cm]{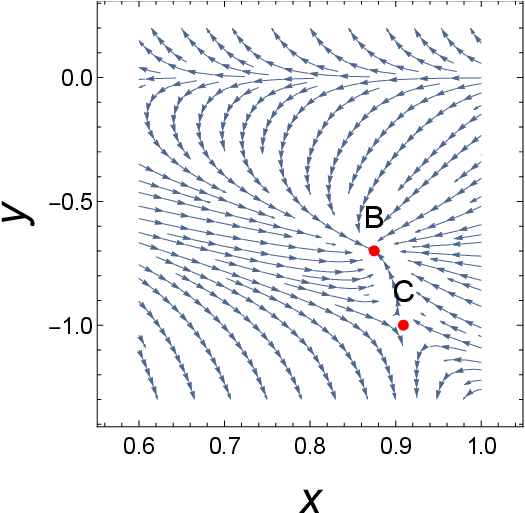}\label{fig:B}}
	\caption{The figure shows the phase portrait of the autonomous system (\ref{autonomous}) for interaction (\ref{Interaction}) in x-y plane for different parameter values.  In panel (a) with the parameter values $\gamma_{m}=0$, $\omega=-0.9$, and $\lambda=0.27$, the scaling solution B is stable attractor. In panel (b), for the parameters $\gamma_{m}=0$, $\omega=-0.8$, and $\lambda=0.1$ the scaling solutions B is stable attractor and C is saddle.}
	\label{phasespace-figure}
\end{figure}
\begin{figure}
	\centering
	\subfigure[]{%
		\includegraphics[width=8cm,height=6cm]{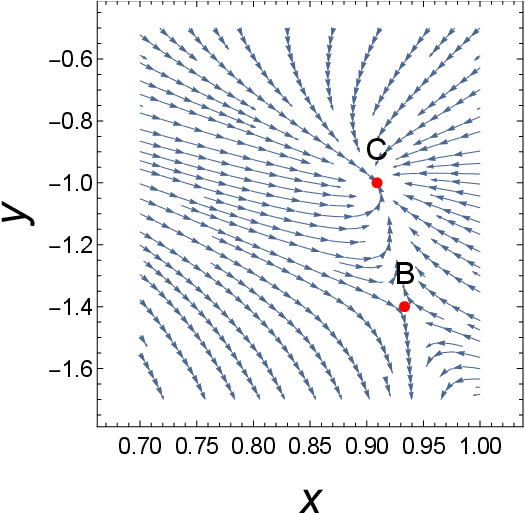}\label{fig:C}}
	\qquad
	\subfigure[]{%
		\includegraphics[width=8cm,height=6cm]{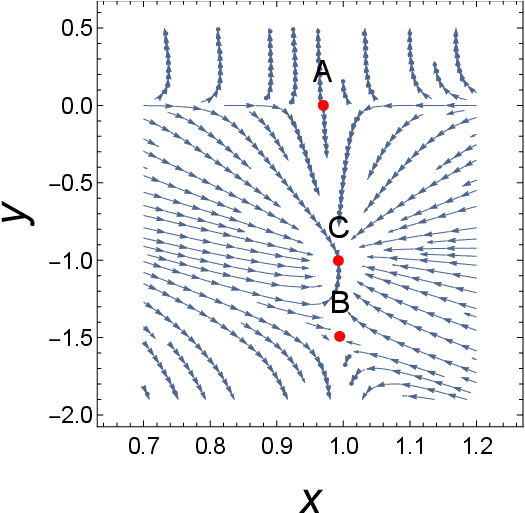}\label{fig:A B C}}
	\caption{The figure shows the phase portrait of the autonomous system (\ref{autonomous}) for interaction (\ref{Interaction}) in x-y plane, where panel (a) with the parameter values $\gamma_{m}=0$, $\omega=1.5$, and $\lambda=0.1$, exhibits the point C is stable attractor and B is unstable saddle and panel (b) for the parameters $\gamma_{m}=\frac{1}{3}$, $\omega=1.5$, and $\lambda=0.01$ shows that the points A and B are saddle-like solutions and point C is stable attractor.}
	\label{phasespace-figure}
\end{figure}


\begin{figure}
	\centering
	\subfigure[]{%
		\includegraphics[width=8cm,height=6cm]{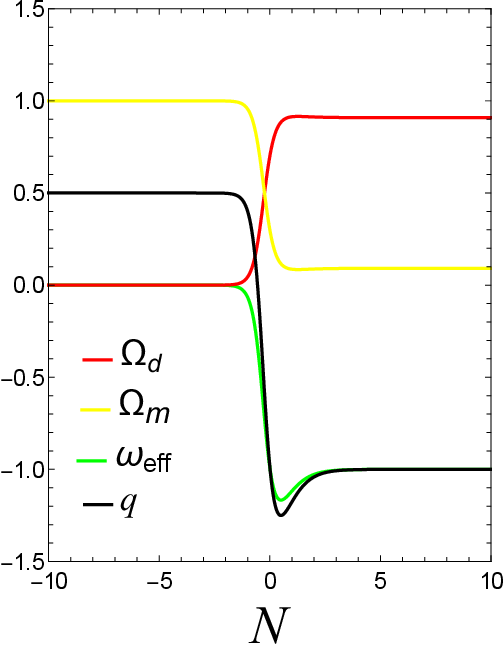}\label{fig:Phantom}}
	\qquad
	\subfigure[]{%
		\includegraphics[width=8cm,height=6cm]{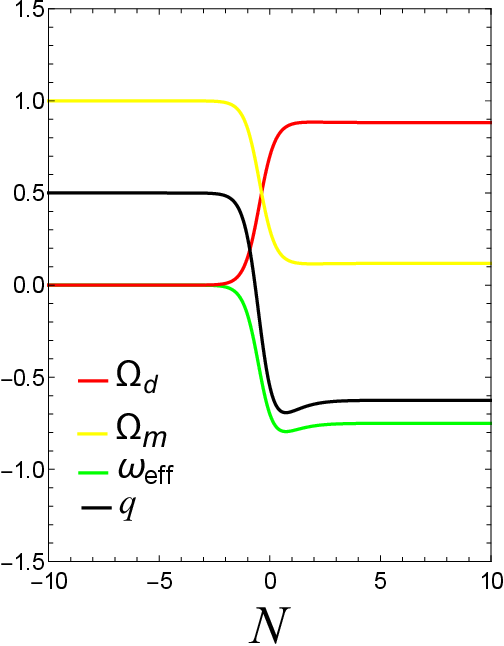}\label{fig:Quintessence}}
	\caption{The figure shows the time evolution of the physical parameters $\Omega_{d}$, $\Omega_{m}$, $\omega_{eff}$ and $q$ for the autonomous system  (\ref{autonomous}). In panel (a) for $\omega=1.5$, $\lambda=0.1$ and $\gamma_{m}=0$, the effective equation of state parameter crosses the phantom divide line. Ultimate fate of the accelerated universe is ($\varLambda$CDM) attracted by cosmological constant like fluid in its future evolution. In (b) the late time attractor is described in quintessence era for the parameter values $\omega=-0.85$, $\lambda=0.1$, and $\gamma_{m}=0$.}
	\label{phasespace-figure-HXY}
\end{figure}

\begin{figure}
	\centering
	\includegraphics[width=9cm,height=7cm]{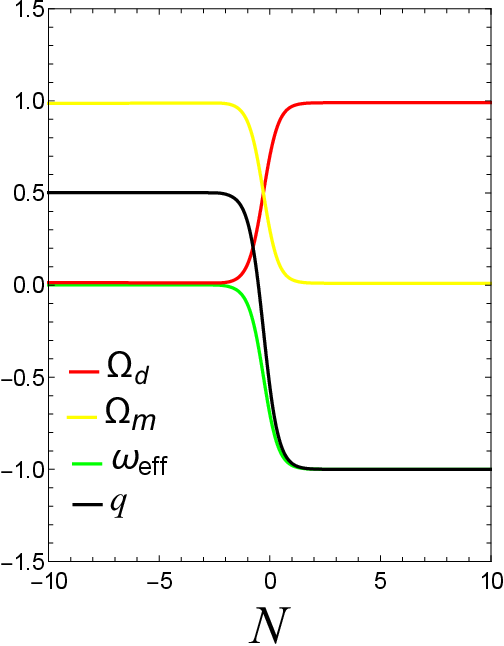}
	\caption{Time evolution of physical parameters shows that late time attractor is ($\varLambda$CDM) achieved by the cosmological constant like fluid for the autonomous system (\ref{autonomous}) with the parameter values $\omega=1.5 $, $\gamma_{m}=0.001$ and $\lambda=0.01$. Late time DE dominated solution is followed by long time matter dominated intermediate phase of the universe.}
	\label{fig:LCDM}
\end{figure}
\subsection{cosmological interpretation:}

From the critical points analysis we observe some  physically interesting phenomena in the evolution of the universe such as for some parameter restrictions, the point A describes the matter dominated phase of the universe which is always decelerating in nature. Here, Umami chaplygin gas behaves as dust like fluid (see figure \ref{fig:A B C} for $\omega=1.5$ and $\lambda=0.01$) and the barotropic fluid is taken as radiation i.e $\gamma_{m}=\frac{1}{3}$).

Late time scaling solutions namely, the critical points B and C are achieved from the dynamical analysis. They are very much interesting from cosmological point of view. Although, the effective equation of state ($\omega_{eff}$) crosses the phantom divide line and even it evolves in phantom regime, but the point B shows unstable saddle like nature there. Therefore, the late time evolution of the universe near the point B cannot be attracted at phantom regime, because the evolution of the universe can only be attracted in quintessence era. For example, the figure \ref{fig:B}, for parameters $\omega=-0.8$, $\lambda=0.1$ and $\gamma_{m}=0$ shows that the point B is the late time stable attractor. In this case, density parameters for DE and DM take values as $\Omega_{d}=0.875$ and $\Omega_{m}=0.125$ respectively. Here, DE behaves as quintessence fluid (since $\gamma_{d}=-0.8$) and the deceleration parameter and the effective equation of state parameter are $q=-0.55$ and $\omega_{eff}=-0.7$ respectively. Interestingly, the point B agrees with the present observations for the parameter values: $\omega=-0.9$, $\lambda=0.27$ and $\gamma_{m}=0$ which are shown in the figure \ref{fig:B_Obsv}, where B is a stable attractor with the cosmological parameters: $\Omega_{d}=0.7$, $\Omega_{m}=0.3$, $q= -0.45$ and $\omega_{eff}= -0.63$. Here, Umami chaplygin gas behaves like quintessence dark fluid $\gamma_{d}=-0.9$. Also, the figure \ref{fig:Quintessence}, for $\omega=-0.85$, $\lambda=0.1$ and $\gamma_{m}=0$ shows that the late time attractors are in quintessence era. Although, the figure \ref{fig:Phantom} for $\omega=1.5$, $\lambda=0.1$, $\gamma_{m}=0$ shows that the point B evolves into phantom regime (as $q=-1.6$ and $\omega_{eff}=-1.4$), but it does not describe the solution at late time (since one of eigenvalues is positive $\mu_{1}=2.4$ and $\mu_{2}=-4.2$ and as a result it is unstable). This is also shown in vector field diagram in figure \ref{fig:C} with the same parameter values. Here, only point C shows late time attractor and this point is evolving in cosmological constant era. Thus, one can conclude that DE-DM scaling solution is represented by the critical point B showing (with some parameter restrictions) both the energy densities of DE and DM scale with the same order of magnitude $r=\frac{\Omega_{m}}{\Omega_{d}}=O(1)$ (satisfying Eqn. (\ref{coincidence})) giving the possible solution of coincidence problem. \\

Another interesting critical point C corresponds to a scaling solution which is an attractor in the phase space. For parameter restriction: $|\omega|>1+\lambda$, the critical point represents the $\varLambda$CDM model of the universe. For an uncoupled (i.e., $\lambda=0$) case, the critical point C describes the late time de Sitter solution which is always accelerating in nature ($q=-1$, $\omega_{eff}=-1$). For the parameter values $\omega=1.5$, $\lambda=0.1$, $\gamma_{m}=0$, the figure \ref{fig:C} and for $\omega=1.5$, $\lambda=0.01$, and $\gamma_{m}=\frac{1}{3}$, the figure \ref{fig:A B C} show that the point C is stable attractor in phase space and the point B is saddle like solution. Also, the evolutions of density parameter of DE ($\Omega_{d}$), DM ($\Omega_{m}$), effecttive equation of state ($\omega_{eff}$) and the decelertion parameter ($q$) with respect to $N$ are shown in the figure (\ref{fig:LCDM}) exhibiting the late time accelerated DE dominated era connecting through a sufficient amount of matter dominated decelerated phase of the universe.

\section{Thermodynamics of universe containing Interacting Umami Chaplygin Gas}
In this section we have followed the work of Gong et al \cite{Gong2007} for studying non-equilibrium thermodynamics of universe bounded by apparent/event horizon. Here, we have considered homogenous, isotropic and flat model of universe as an open thermodynamical system. Following Gong et al, we have considered an irreversible process: an energy flux going through the horizon of universe. Due to this irreversibility, an internal entropy production $dS_i$ should be taken into account according to nonequilibrium thermodynamics.\\

The associated theory of irreversible thermodynamics is as follows:\\
In non-equilibrium thermodynamics, Clausius law is replaced by the following entropy balance law:\\
\begin{equation*}
  dS=\frac{\delta Q}{T}+dS_{i}
\end{equation*}
where the first term on the right hand side is the usual exchange of entropy between the system and its surroundings and the second term arises from the internal production process due to the associated irreversible process. Here $\delta Q$ is classically referred to as compensated heat i.e the heat transferred between the system and its surroundings while $dS_{i}$ is called uncompensated heat indicating the amount of entropy associated with the heat which is intrinsic to the system when
it undergoes an irreversible process. In particular,\\
$
dS_{i}=
\begin{cases}
>0,~~ \textrm{for irreversible process}\\
0, ~~ \textrm{for reversible process}
\end{cases}
$\\
Then,in general, the change of entropy of a system can be then written as
\begin{equation}
	dS=dS_{e}+dS_{i}\label{2}
\end{equation}
where $dS_{e}=\frac{dQ}{T}$.
Now, suppose $\sigma$ represents the internal entropy source production density and $\overrightarrow{J_{s}}$ denotes entropy flow density and then assuming local equilibrium one can write
\begin{equation}
	\frac{dS_{e}}{dt}=-\int_{\Sigma}\overrightarrow{J_{s}}.d\overrightarrow{\Sigma}\label{3}
\end{equation}
and
\begin{equation}
	\frac{dS_{i}}{dt}=\int_{V}\sigma dV\label{4}
\end{equation}
where $d\overrightarrow{\Sigma}$ is the oriented surface element and $V$ is the volume bounded by the horizon and $\Sigma$ denotes the surface of the horizon. Though different heat mechanism namely convection, heat conduction, diffusion and other processes may caused the entropy flow $\overrightarrow{J_{s}}$ and internal entropy source production $\sigma$, but for simplicity we consider only the heat conduction as the dominant contributor. Thus we have
\begin{equation}
	\overrightarrow{J_{s}}=\frac{\overrightarrow{J_{q}}}{T}\label{5}
\end{equation}
\begin{equation}
	\sigma=\overrightarrow{J_{q}}.\nabla\left(\frac{1}{T}\right)\label{6}
\end{equation}
with $\overrightarrow{J_{q}}$ the heat current due to conduction and $T$, the temperature of the system. Now it is convenient to assume $\overrightarrow{J_{q}}$ and $\sigma$ to be uniform across the surface of the horizon and the volume bounded by the horizon respectively. Then we have from equation (\ref{3}) using (\ref{5})
\begin{equation}
	\frac{dS_{e}}{dt}=|\overrightarrow{J_{q}}|\frac{A}{T}\label{7}
\end{equation}
and also from equation (\ref{4})
\begin{equation}
	\frac{dS_{i}}{dt}=\sigma.V\label{8}
\end{equation}
where $A$ and $V$ are the surface area of the horizon and the amount of volume bounded by the horizon respectively.\\
As there is a spontaneous heat flow between the horizon and the dark energy fluid due to non-equilibrium thermodynamics, so the heat current $\overrightarrow{J_{q}}$ can be measured by the temperature gradient using the Fourier law as
\begin{equation}
\overrightarrow{J_{q}}=-\lambda^{\prime}\nabla(T)\label{9}
\end{equation}
where $\lambda^{\prime}$~ is the thermal conductivity.\\\\
Now, for a Bekenstein system entropy of horizon is given by [] $S_{\sum}=\frac{A}{4}$. Then considering the universe to be a Bekenstein system,then entropy of the horizon  and temperature becomes
\begin{eqnarray}
 S_{\sum}&=&\pi R_{\sum}^{2}\\
 T&=& \frac{1}{2\pi R_{\sum}}\label{10}
\end{eqnarray}

where $R_{\sum}$ is the radius of bounding horizon.\\
Using the above Bekenstein entropy (\ref{10}) in eq.(\ref{7}) we obtain the heat current as
\begin{equation}
|\overrightarrow{J_{q}}|=\frac{\dot{R_{\sum}}T}{2R_{\sum}}\label{11}
\end{equation}
Also, eliminating temperature gradient and heat current from the equations (\ref{6}),(\ref{9}) and (\ref{11}) we get
\begin{equation}
\sigma=\frac{\dot{R_{\sum}}^{2}}{4R_{\sum}^{2}\lambda}\label{12}
\end{equation}
Thus the time variation of the entropy due to entropy production can be obtained using equation (\ref{8})and (\ref{12}) as
\begin{equation}
	\frac{dS_{i}}{dt}=\frac{\pi R_{\sum}\dot{R_{\sum}}^{2}}{3\lambda}\label{13}
\end{equation}
Hence the total entropy change of the apparent horizon can be expressed as
\begin{equation}
	\frac{dS_{T}}{dt}=2\pi R_{\sum}\dot{R_{\sum}}\left(1+\frac{\dot{R_{\sum}}}{6\lambda}\right)\label{14}
\end{equation}
which shows the dependence of the entropy of the horizon on the non-equilibrium factor $\lambda$.\\\\

Now, we shall examine our current model in the light of above prescribed general theory of irreversible thermodynamics. With this purpose, for simplicity, we have assumed dark matter to be in the form of pressureless dust $\left(i.e~ p_{m}=0\right)$. Then we solve energy conservation equations (\ref{continuity-DE}) and (\ref{continuity-matter}) for $\rho_d$ and $\rho_m$. It is to be mentioned that it is not possible to express $\rho_d$  explicitly in terms of scale factor $a$ when we solve equation (\ref{continuity-DE}) using expression for $p_d$ given by equation (\ref{1}). Therefore, we take binomial expansion in the expression of $p_d$ in (\ref{1}) into consideration up to first order in ${\rho_{d}}^{2}$ and then solving the equation (\ref{continuity-DE}) we get

\begin{equation}
	\rho_{d}=\rho_{0}\left(1-B_{s}+B_{s}a^{-\mu}\right)^{-1/2}\label{17}
\end{equation}
Putting the above expression for $\rho_{d}$ in (\ref{continuity-matter}) we obtain
\begin{equation}
	\rho_{m}=\frac{\lambda\rho_{0}}{\sqrt{1-B_{S}}} H_{2}F_{1}\left[\frac{1}{2},-\frac{3}{\mu},\frac{-3+\mu}{\mu},\frac{-B_{s}a^\mu}{1-B_{s}}\right] \label{18}
\end{equation}
where
\begin{equation}
\mu = 1+\lambda-\frac{1}{|\omega|}\label{26}
\end{equation}
\begin{equation}
B_{s}= 1+\frac{\rho_{0}^2}{\mu|A|}\label{27}
\end{equation}
$\rho_{0}$ being the energy density of DE at present time.
The analytic form of  equation of state parameter (EoS) $\gamma_d$ for Umami Chaplygin gas then reads as
\begin{equation}
	\gamma_d=-\frac{1}{\frac{1}{|\omega|}+\frac{{\rho_{0}}^2}{|A|}\left(1-B_{s}+B_{s}a^{-\mu}\right)^-1}\label{19}
\end{equation}
In order to avoid phantom era of DE, we restrict $\gamma_d$ to lie within -1 and $-1/3$ i.e  $ -1<\gamma_d<-1/3$
which renders bounds on $\rho_{0}$ as
\begin{equation}
	|A|\left(1-\frac{1}{|\omega|}\right)<{\rho_{0}}^2<|A|\left(3-\frac{1}{|\omega|}\right)\label{20}
\end{equation}
\subsection*{\textbf{Validity of Generalized Second Law of Thermodynamics}}
Here the validity of generalized second law of thermodynamics has been studied for the universe filled with interacting DE and DM. Also,a series of recent observations involving galaxies and cluster of galaxies as well as the cosmic microwave background \cite{Bennet2003} reveal that present universe is close to flat. So it is legitimate to consider the universe to be flat in this paper. In two separate cases when universe is bounded by apparent horizon and event horizon respectively,the GSLT has been examined in the context of irreversible thermodynamics.
\subsection*{\textbf{Case-I : Apparent Horizon}}
In flat universe radius of apparent horizon is $R_{A}=\frac{1}{H}$.\\
From Friedmann equations for flat universe we can obtain $R_{A}$ and $\dot{R}_{A}$~ using the expression for $\rho_{d}$ and $\rho_{m}$ in equations (\ref{17}) and (\ref{18})as
\begin{equation}
	R_{A}=\sqrt{\frac{3}{8\pi\rho_{0}}}\left(l+\frac{\lambda m}{\sqrt{1-B_{s}}}\right)^{-1/2}\label{21}
\end{equation}
\begin{equation}
	\dot{R}_{A}=\frac{3}{2}\left[1-\left(\frac{1}{|\omega|}+\frac{\rho_{0}^2}{\mu|A|}\right)^{-1}\left(1+\frac{\lambda m}{l\sqrt{1-B_{s}}}\right)^{-1}\right] \label{22}
\end{equation}
where
\begin{eqnarray*}
	l &=& \left(1-B_{s}+B_{s}a^{-\mu}\right)^{-1/2} \\
	m &=& H_{2}F_{1}\left[\frac{1}{2},-\frac{3}{\mu},\frac{-3+\mu}{\mu},\frac{-B_{s}a^\mu}{1-B_{s}}\right]
\end{eqnarray*}
However, in order to get real value of $R_{A}$ we must have:
\begin{eqnarray}
	l+\frac{\lambda m}{\sqrt{1-B_{s}}} &>& 0 \nonumber\\
	B_{s} &<& 1\label{23}
\end{eqnarray}
Now, the condition given by (\ref{23}) imposes further constraints as follows:
\begin{eqnarray}
	B_{s} &=&1+\frac{\rho_{0}^2}{\mu|A|}<1 \nonumber \\
	\Rightarrow \frac{\rho_{0}^2}{\mu|A|}&<& 0\nonumber\\
	\mu & < & 0\label{24}
\end{eqnarray}
Once we obtain that $\mu<0$, it follows that $\omega$ should satisfy the following inequality:
\begin{equation}
	|\omega| < \frac{1}{1+\lambda}\label{25}\\
\end{equation}
i.e acceptable values of  $\omega$ will lie in a range that will depend on the interaction parameter $\lambda$.\\\\
However, GSLT is satisfied if $\frac{dS_{T}}{dt}\geq 0$. From equation (\ref{14}) it is evident that since $R_{A}>0$ always, GSLT will be valid if one of the following conditions are satisfied:
\begin{eqnarray*}
	(i) \dot{R}_{A}&>& 0 \\
	(ii) \dot{R}_{A} &<& -6\lambda^{\prime}
\end{eqnarray*}
Now, from the expression of $\dot{R}_{A}$ given by equation (\ref{22}), it is not possible analytically to decide whether $\dot{R}_{A}>0$ or $\dot{R}_{A}<0$. Therefore graphically we have tried to check about the sign of $\dot{R}_{A}$. In order to do this we have considered the best fit values of the model parameters $\omega$ and $A$ as suggested by R. Lazkoz in \cite{Lazkoz2019}. In particular we chose $\omega=-0.85$ and $\overline{A}=0.5$ .\\
Here $\overline{A}=\frac{A}{{\rho_{c,0}}^2}$ and  $\rho_{c,0}=\frac{3{H_{0}}2}{8\pi G}$ is the critical density of the universe at present time.\\

Given the value of $\omega$, we have to choose $\lambda$ in such a way that it satisfies the restriction (\ref{25})i.e
\begin{eqnarray*}
	\lambda &<& \frac{1}{|\omega|}-1
\end{eqnarray*}
Since we have chosen $\omega=-0.85$, therefore chosen $\lambda$ will always be less than 0.18. Also (\ref{27})can be rewritten as
\begin{eqnarray}
	B_{s}&=&1+\frac{\rho_{0}^2}{\mu|A|}\nonumber\\
	~&=&1+\frac{\rho_{0}^2/{\rho_{c,0}}^2}{\mu|A|/{\rho_{c,0}}^2}\nonumber\\
	~&=&1+\frac{\Omega_{0}^2}{\mu|\overline{A}|}
\end{eqnarray}
With present value of dark energy density parameter $\Omega_{0}=0.623$ and $\mu$ computed from equation (\ref{26})and best fit values of model parameters $\omega=-0.85$ and $\overline{A}=0.5$ we can calculate values of $B_{s}$ for different $\lambda$.\\

With the above choice of parameters and constants in figure-1 we have plotted $\dot{R}_{A}$ against the scale factor for different $\lambda$.\\
\begin{figure}[tb]
\centering
	\includegraphics[width= 1.0\columnwidth]{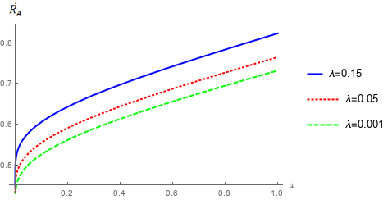}
	\caption{Graphical representation of~ $\dot{R}_{A}$ against $a$ ~for ~$\omega=-0.85,\overline{A}=0.5$, $\Omega_{0}=0.623$~~and ~different $\lambda$}
	\label{fig:1}
\end{figure}
From figure 1 we see that $\dot{R}_{A}>0$ and therefore it can be conjectured that GSLT will be valid for best fit values of the model parameters and for some restricted range of values of interaction parameter $\lambda$.
\subsection*{\textbf{Case-II: Event Horizon}}
By definition the radius of event horizon is given by
\begin{equation}
R_{E}=a\int_{t}^{\infty}\frac{dt}{a(t)}\label{28}
\end{equation}
Therefore the time derivation of ~$R_{E}$~ is given by
\begin{equation}
\dot{R_{E}}=HR_{E}-1\label{29}
\end{equation}
Now, let consider the UCG to satisfy holographic dark energy (HDE) principle. According to HDE model using effective quantum field theory the energy density is given by \cite{Akarsu2010,Horava2000}~~$\rho_{d}=\frac{3c^{2}}{R_{E}^{2}}$\\
where~ `c'~ is a dimensionless parameter which may be estimated from observation \cite{Huang2004,Chang2006,Zhang2005,Saridakis2007} and the radius of the event horizon is chosen as the IR cut-off length to obtain correct equation of state and the desired accelerating universe \cite{Li2004}. So one can write ~$R_{E}$~ as~~~$R_{E}=\frac{c}{\sqrt{\Omega_{d}}H}$\\\\
where~ $\Omega_{d}=\frac{8\pi\rho_{d}}{3H^{2}}$~ is the density parameter.\\\\
Now the equation of state parameter of the interacting holographic DE has the form
\begin{equation}
\omega_{D}=-\frac{1}{3}-\frac{2\sqrt{\Omega_{D}}}{3c}-\lambda\label{30}
\end{equation}
If one makes a correspondence between UCG and HDE, from (\ref{19}) and (\ref{30}) it can be obtained that
\begin{equation}
\sqrt{\Omega_{d}}=\frac{3c}{2}\left[\frac{1}{\frac{1}{|\omega|}+\frac{\rho_{0}^2}{|A|\left(1-B_{s}+B_{s}a^{-\mu}\right)}}-\left(\lambda+\frac{1}{3}\right)\right]\label{31}
\end{equation}
From (\ref{13}) the time variation of total entropy for a universe bounded by event horizon and filled with interacting HDE can be derived as
\begin{eqnarray}
\frac{dS_{T}}{dt}=2\pi\frac{cH^{-1}}{\sqrt{\Omega_{d}}}\left(\frac{c}{\sqrt{\Omega_{d}}}-1\right)
\left[1+\frac{c}{6\lambda \sqrt{\Omega_{d}}}\left(1-\frac{\sqrt{\Omega_{d}}}{c}\right)\right]\label{32}
\end{eqnarray}
The above expression shows that if  ~$\sqrt{\Omega_{d}}< c$, then the validity of GSLT is guaranteed.
This restriction imposes the following restriction at present time, i.e for $a=1$ :
\begin{equation}
\rho_{0}> \sqrt{|A|\left(\frac{1}{1+\lambda}-\frac{1}{|\omega|}\right)}\label{33}
\end{equation}
and
\begin{equation}
|\omega|>1+\lambda\label{34}
\end{equation}

But for the above restriction (\ref{34}), expression for $\rho_m$ and Hubble parameter $H$ do not possess real value and hence this restriction is not plausible. Therefore, we can conjecture that GSLT is not valid on event horizon.




\section{Summary and concluding remarks}
\label{summary}

In this work, authors have investigated the thermodynamic as well as dynamical behaviour of the universe in the context of interacting Umami chaplygin gas(UCG) Dark Energy Model. In this section, first we discuss the results obtained from dynamic and themodynamic analysis separately and finally combining these two fold results. Also, we discuss the physically permissible limit of model parameters, which are relevant both from dynamic as well as thermodynamic point of view.\\

Dynamical systems analysis of interacting Umami chaplygin gas has been studied in the background of spatially flat FLRW model of the universe where Umami chaplygin gas acted as dark energy interacting with DM which is in the form of perfect fluid with barotropic equation of state $p_m=\gamma_{m}\rho_{m}$. Cosmological evolution equations have been transformed into an autonomous system of ODEs by suitable choices of dimensionless dynamical variables normalized over Hubble scale. We have obtained the critical points by equating the right part of the system to zero. The nature of critical points are found by giving a small perturbation around the critical points. Stability of the critical points are obtained by eigenvalues of the linearized Jacobian matrix evaluated at the critical points. \\

From the dynamical analysis, several physically interesting critical points are achieved. Among them, the critical point A represents completely Umami Chaplygin gas (DE) dominated solution but Umami Chaplygin gas behaves here like a dust fluid (as dark matter) and also, there exists decelerating phase of the universe near the critical point. Being unstable, the point does not represent the solution at late time. So, the point has the nature to describe dust dominated decelerated intermediate phase of the universe.\\

Next, the late time accelerated DE-DM scaling attractor is represented by the critical point B which has the evolutionary phenomenon in phantom regime (i.e., $\omega_{eff}<-1$), but the point is not of much physically interesting in that region due to its unstable nature there (see figure \ref{fig:Phantom}, \ref{fig:C}). However, the point will become physically interesting only when the parameters satisfy the following restriction: $|\omega|<1+\lambda$. In this parameter region, the point behaves as late time stable attractor. As for example, the figure \ref{fig:B_Obsv} shows the critical point B can succesfully predict the present observed phenomena of the universe. Moreover, in that restrcition, the critical point also behaves as a scaling attractor in quintessence era, where the expansion of the universe is always accelerating. In this connection, the figure \ref{fig:Quintessence} exhibits the characteristics of late time evolution of the universe in quintessence era. \\

Finally, the critical point C becomes a cosmologically viable solution at late time only for the parameter restriction: $|\omega|>1+\lambda$. Also, the critical point corresponds to a scaling solution which is accelerating always. From the time evolution of cosmological parameters (in figure \ref{fig:LCDM}), one can conclude that accelerated de Sitter era of the universe is achieved at late time connecting through an intermediate matter dominated decelerated phase of the universe. This phenemenon is ineresting in an interacting scenario. Here, the ultimate fate of the universe is described by a cosmological constant fluid.\\

Therefore, from the dynamical study of the model one can conjecture that the Umami Chaplygin gas can behave as dust (DM) exhibiting the matter (dust) dominated intermediate phase of the universe (presented by the critical point A). Also, the Umami Chaplygin gas can behave as quintessence which produces the late time accelerated scaling solution (presented by the critical point B) alleviating the coincidence problem successfully and the universe evolves in quintessence era in this case. Finally, the model exhibits late time evolution of the accelerated phase of the universe (presented by the critical point C) attracted by the cosmological constant when Umami Chaplygin gas behaves as phantom fluid.\\

For thermodynamical analysis standard(Eckart)theory of irreversible thermodynamics has been applied with apparent/event horizon as bounding horizon. In particular the flat universe is considered to be composed of two dark sectors- dark energy and dark matter. UCG has been chosen to play the role of dark energy while pressureless dust serves to dark matter. Generalized second law of thermodynamics (GSLT) has been studied for both apparent and event horizon in two separate cases. The results points out to some interesting facts. Firstly, it shows that the EoS parameter $\gamma_d$ does not cross the phantom divide if the present value of dark energy density $\rho_{0}$ lies in a interval and the bounds of the interval depend on both the model parameters $A$ and $\omega$. Secondly, for having realistic value of apparent horizon radius, model parameter $\omega$ and the interaction  parameter$\lambda$ should be related as $|\omega|<\frac{1}{1+\lambda}$. Thirdly, for GSLT to be valid it is seen that time derivative of $R_{A}$ should be either positive or $<-6\lambda^{\prime}$ where $\lambda^{\prime}$ is the non-equilibrium factor (Thermal conductivity). Due to complex form of $\dot{R}_{A}$, it has analytically not been possible to draw any conclusion. Therefore graphically with some particular values of model parameters as suggested by R. Lazkoz in \cite{Lazkoz2019}, it is shown that $\dot{R}_{A}$ is positive and therefore GSLT is valid on apparent horizon if parameters are suitably chosen. It is to be noted that the best fit values of parameters predicted in \cite{Lazkoz2019} satisfy the restriction $|\omega|<\frac{1}{1+\lambda}<1+\lambda$. Lastly, it is shown that GSLT is not valid on event horizon where UCG was considered to be holographic dark energy interacting with dark matter as it imposes an unrealistic constraint for the validity of GSLT.\\

 From the results obtained from dynamical and thermodynamic analysis, we get an clear idea about the viable range of the model parameter $\omega$. We see that feasible limit of $\omega$ depends explicitly on the value of interaction (coupling) parameter $\lambda$. In particular, for $|\omega|< 1+\lambda$, it is revealed from dynamical analysis that the current model can describe late time acceleration of universe and can alleviate cosmic coincidence problem. This range of $\omega$ is also important from thermodynamic point of view as generalized second law of thermodynamics (GSLT) is satisfied on apparent horizon when $\omega$ lies in this range, to be specific when $|\omega|<\frac{1}{1+\lambda}$\\
  On the other hand, for $|\omega|>1+\lambda$, dynamical analysis shows that the current model correctly explains evolution of universe from matter dominated deceleration phase to dark energy dominated accelerating phase solving coincidence problem as well. It actually describes $\Lambda$CDM cosmology within this range of $\omega$ which is observationally very relevant. But from thermodynamic point of view this range of $\omega$ is not physically acceptable. In this regard, it is to be mentioned that exact analytic expression for dark energy density and dark matter density were not possible to calculate from energy conservation equations and so we worked with approximate solutions. These approximate solutions can give us idea of only near exact scenario. Therefore, future challenge lies in developing suitable technique for finding exact solutions of energy densities of interacting Umami fluid and dark matter and to explore thermodynamic consequences when $|\omega|>1+\lambda$.\\
  Also, the current model could not predict anything about the other model parameter $A$. A further extensive investigation is needed to obtain complete information about both the parameters.

\section*{Acknowledgments}
The authors are grateful to the anonymous referee for his critical review and valuable comments as it helped the authors to improve the quality of the manuscript significantly.

\end{document}